\documentclass[10pt]{iopart}
\usepackage{iopams,bm,cite,color,graphicx,mathptmx,url}

\newcommand{\op}[1]{\hat{#1}}

\newcommand{\openone}{\leavevmode\hbox{\small1\normalsize\kern-.33em1}}

\eqnobysec

\begin{document}

\title{Quantum versus classical polarization states: when multipoles count}

\author{  L~L~S\'anchez-Soto$^{1,2,3}$, A~B~Klimov$^{4}$, 
P~de la Hoz$^{3}$  and G~Leuchs$^{1,2}$}

\address{$^{1}$ Max-Planck-Institut f\"ur die Physik des Lichts,
  G\"{u}nther-Scharowsky-Stra{\ss}e 1, Bau 24, 91058 Erlangen,
  Germany}

\address{$^{2}$ Institut f\"ur Optik, Information und Photonik,
  Staudtstra{\ss}e 7, 91058 Erlangen, Germany}

\address{$^{3}$ Departamento de \'Optica, Facultad de F\'{\i}sica,
  Universidad Complutense, 28040~Madrid, Spain}

\address{$^{4}$ Departamento de F\'{\i}sica, Universidad de
  Guadalajara, 44420~Guadalajara, Jalisco, Mexico}

\date{\today}

\begin{abstract}
  We advocate for a simple multipole expansion of the polarization
  density matrix. The resulting multipoles are used to construct
  \textit{bona fide} quasiprobability distributions that appear
  as a sum of successive moments of the Stokes variables;
  the first one corresponding to the classical picture on the
  Poincar\'e sphere. We employ the particular case of the $Q$ function
  to formulate a whole hierarchy of measures that properly assess
  higher-order polarization correlations.
\end{abstract}


\section{Introduction}

Polarization is a fundamental property of light that has received a
lot of attention over the years~\cite{Brosseau:1998lr}. As
polarization is a robust characteristic, relatively simple to
manipulate without inducing more than marginal losses, it is not
surprising that many experiments at the forefront of quantum optics
involve this observable~\cite{Mandel:1995qy}.

In classical optics, polarization is elegantly visualized using the
Poincar\'e sphere and is determined by the Stokes parameters.  These
are measurable quantities that allow for a classification of the
states according to a degree of polarization. Furthermore, the
formalism can be extended to the quantum domain, where the Stokes
parameters become the mean values of the Stokes
operators~\cite{Luis:2000ys}.

The classical degree of polarization is just the length of the
Stokes vector. This provides a very intuitive picture, but for intricate
fields it has serious drawbacks. Indeed, this classical quantity does
not distinguish between states having remarkably different
polarization properties~\cite{Tsegaye:2000mz}. In particular, it can
be zero for light that cannot be regarded as unpolarized, giving rise
to the so-called \textit{hidden polarization}~\cite{Klyshko:1992wd}.  All
these flaws have prompted some alternative measures~\cite{Luis:2002ul,
Legre:2003qf,Saastamoinen:2004lq,Picozzi:2004vn,Ellis:2005cr,
Luis:2005qf,Sehat:2005wd,Klimov:2005kl,Refregier:2005uq,
Refregier:2006fj,Luis:2007kx,Bjork:2010rt,Klimov:2010uq,Qian:2011kx}.

We adhere to the viewpoint that the Stokes measurements ought to be
the basic building blocks for any practical approach to
polarization. Actually, the aforesaid problems with the classical
degree are due to its definition in terms exclusively of first-order
moments of the Stokes variables.  This may be sufficient for most
classical situations, but for quantum fields higher-order correlations
might be crucial.

Our goal in this paper is to advance a practical solution to these
hurdles.  From coherence theory, we learn that a complete description of
interference phenomena involves a hierarchy of correlation functions,
with classical behavior represented by the first one of those.  In the
same spirit, we propose to go beyond the first-order description and
look for a way to systematically assess higher-order polarization
correlations.

For that purpose, we borrow basic ideas from the standard theory of
SU(2) quasidistributions~\cite{Varilly:1989ud}, but we reinterpret
them in terms of multipoles that contain sequential moments of the
Stokes variables. The dipole, being just the first-order moment, can
be identified with the classical picture, whereas the other multipoles
account for higher-order correlations.  Finally, we illustrate how the
particular instance of the SU(2) $Q$ function can be used as an
efficient measure for the quantitative assessment of those
fluctuations.

\section{Polarization structure of quantum fields}

We start with a brief survey of the basic ingredients involved in
a proper description of quantum polarization. We assume a
monochromatic plane wave, propagating in the $z$ direction, so its
electric field lies in the $xy$ plane. We are thus effectively dealing
with a two-mode field that can be characterized by two complex
amplitude operators, denoted by $\op{a}_{\mathrm{H}}$ and
$\op{a}_{\mathrm{V}}$, where the subscripts H and V indicate label
horizontal and vertical polarization modes. These operators obey the
commutation rules $ [\op{a}_{\lambda}, \op{a}_{\mu}^\dagger ] =
\delta_{\lambda \mu}$, with $\lambda , \mu \in \{ \mathrm{H},
\mathrm{V} \}$.

The use of the Schwinger representation~\cite{Schwinger:1965kx}
\begin{equation}
  \label{Stokop}
  \fl
  \textstyle   
  \op{S}_{1} = \frac{1}{2} 
  ( \op{a}_{\mathrm{H}}^{\dagger}   \op{a}_{\mathrm{V}} + 
  \op{a}_{\mathrm{V}}^{\dagger}  \op{a}_{\mathrm{H}} ) \, ,
  \qquad
  \op{S}_{2} =  \frac{i}{2} 
  ( \op{a}_{\mathrm{H}} \op{a}_{\mathrm{V}}^{\dagger}  - 
  \op{a}_{\mathrm{H}}^{\dagger}  \op{a}_{\mathrm{V}}) \, ,
  \qquad
  \op{S}_{3}  = \frac{1}{2} 
  ( \op{a}_{\mathrm{H}}^{\dagger} \op{a}_{\mathrm{H}} -  
  \op{a}_{\mathrm{V}}^{\dagger} \op{a}_{\mathrm{V}} ) \, , 
\end{equation}
together with the total number operator $ \op{N} =
\op{a}_{\mathrm{H}}^{\dagger} \op{a}_{\mathrm{H}} +
\op{a}_{\mathrm{V}}^{\dagger} \op{a}_{\mathrm{V}}$, will prove very
convenient in what follows. In fact, the average of $\mathbf{\op{S}} =
( \op{S}_{1}, \op{S}_{2}, \op{S}_{3})$ coincides (except for an
unimportant factor 1/2) with the classical Stokes
vector~\cite{Luis:2000ys}. Such a numerical factor is inserted 
to guarantee that $\{ \op{S}_{k} \}$ satisfy the commutation
relations of the su(2) algebra
\begin{equation}
  \label{crsu2}
  [ \op{S}_{k}, \op{S}_\ell] = 
  i \epsilon_{k \ell m} \,  \op{S}_{m}  \, , 
\qquad
  [\op{N}, \op{S}_{k} ] = 0 \, ,
\end{equation}
where the Latin indices run over $\{ 1, 2, 3 \}$ and 
$\epsilon_{k \ell  m}$ is the Levi-Civita fully antisymmetric tensor.  
This noncommutability precludes the simultaneous exact 
measurement of the physical quantities they represent, which is 
expressed by the uncertainty relation
\begin{equation}
  \label{eq:ursu2}
  \Delta^{2} \op{\mathbf{S}}  = 
  \Delta^{2} \op{S}_{1}  + \Delta^{2} \op{S}_{2}  + \Delta^{2} \op{S}_{3}  
  \geq \langle \op{N} \rangle /2 \, ,
\end{equation}
$\Delta^{2} \op{S}_{k} = \langle \op{S}_{k}^{2} \rangle -
\langle \op{S}_{k} \rangle^{2}$ standing for the variance.

In classical optics, the states of definite polarization are specified
by the constraint $\langle \op{\mathbf{S}} \rangle^{2} = \langle
\op{N}/2 \rangle^{2}$. Since the intensity is there a nonfluctuating
quantity, in the three-dimensional space of the Stokes parameters this
define a sphere with radius equal to the intensity: the Poincar\'e
sphere.  In contradistinction, in quantum optics we have that
$\op{\mathbf{S}}^{2} = S(S+1) \openone$, with the angular momentum
being $S = N/2$ and, as fluctuations in the number of photons are
unavoidable, we are forced to work in the three-dimensional Poincar\'e
space that can be regarded as a set of nested spheres with radii
proportional to the different photon numbers that contribute to the
state.

As our final remark, we stress that the second equation in
(\ref{crsu2}) prompts to address each subspace with fixed number of
photons $N$ separately.  To bring this point out more clearly, it is
advantageous to relabel the standard two-mode Fock basis
$|n_{\mathrm{H}}, n_{\mathrm{V}} \rangle$ in the form
\begin{equation}
  \label{invsub}
  | S, m \rangle = | n_{\mathrm{H}} = S+m, n_{\mathrm{V}}  = S- m \rangle \, ,
\end{equation}
so that $S = N/2$ and $m = (n_{H} - n_{V})/2$. For each fixed $S$, $m$
runs from $-S$ to $S$ and the states (\ref{invsub}) span a
$(2S+1)$-dimensional subspace wherein $\op{\mathbf{S}}$ act in the
standard way.

\section{The polarization sector}

For any arbitrary function of the Stokes operators $f(\op{\mathbf{S}}
) $, we have $ [ f(\op{\mathbf{S}}), \op{N} ] = 0$ as well, so the
matrix elements of the density matrix $\op{\varrho}$ connecting
subspaces with different photon numbers do not contribute to $\langle
f(\op{\mathbf{S}}) \rangle$. This translates the fact that
polarization and intensity are, in principle, independent concepts: in
classical optics the form of the ellipse traced out by the electric
field (polarization) does not depend on its size (intensity).

In other words, the only accessible information from $\op{\varrho}$ is
its polarization sector~\cite{Karassiov:1993lq,Raymer:2000zt,
  Karassiov:2004xw,Marquardt:2007bh,Muller:2012ys}, which is specified
by the block-diagonal form
\begin{equation}
  \label{eq:decrj}
  \op{\varrho}_{\mathrm{pol}} = \bigoplus_{S}  \op{\varrho}^{(S)} 
\end{equation}
where $\op{\varrho}^{(S)}$ is the reduced density matrix in the $S$th
subspace ($S$ runs over all the possible photon numbers, i. e., $ S =
1/2, 1, \ldots$). Any $\op{\varrho}$ and its associated block-diagonal
form $ \op{\varrho}_{\mathrm{pol}} $ cannot be distinguished in
polarization measurements; accordingly, we drop henceforth the
subscript pol.

To go ahead, we resort to the standard SU(2)
machinery~\cite{Blum:1981ya} and expand each $\op{\varrho}^{(S)} $ as
\begin{equation}
  \label{rho1}
  \op{\varrho}^{(S)} =  \sum_{K= 0}^{2S} \sum_{q=-K}^{K}  
  \varrho_{Kq}^{(S)} \,   \op{T}_{Kq}^{(S)} \, ,
\end{equation}
where the irreducible tensor operators $\op{T}_{Kq}^{(S)}$ (please,
note carefully that the index $K$ takes only integer values)
read~\cite{Varshalovich:1988ct}
\begin{equation}
  \label{Tensor} 
  \op{T}_{Kq}^{(S)} = \sqrt{\frac{2 K +1}{2 S +1}} 
  \sum_{m,  m^{\prime}= -S}^{S} C_{Sm, Kq}^{Sm^{\prime}} \, 
  |  S , m^\prime \rangle \langle S, m | \, ,
\end{equation}
and the expansion coefficients
\begin{equation}
  \varrho_{Kq}^{(S)} =  \Tr [ \op{\varrho}^{(S)} \,
  T_{Kq}^{(S) \, \dagger} ]
\end{equation}
are known as state multipoles and contain all the information about
the state.  The quantities $ C_{Sm, Kq}^{Sm^{\prime}}$ are the
Clebsch-Gordan coefficients that couple a spin $S$ and a spin $K$
to a total spin $S$ and vanish unless the usual angular momentum
coupling rules are satisfied, namely
\begin{equation}
  \label{eq:crul}
  0 \le K \le 2S \, , 
\qquad
- K \le q  \le K \, .
\end{equation}

The operators $\op{T}_{Kq}^{(S)}$ are quite a convenient tool for they
have the proper transformation properties under rotations and besides
fulfill
\begin{equation}
  \Tr  [ T_{K q}^{(S) \, \dagger} \, T_{K^{\prime} q^{\prime}}^{(S)}  ]  =
  \delta_{K K^{\prime}}  \delta_{q q^{\prime}} \, , 
\end{equation}
so, they indeed constitute the most suitable orthonormal basis for the
problem at hand.  Although the definition of $\op{T}_{K q}^{(S)}$ in
(\ref{Tensor}) might look a bit unfriendly, the essential observation
for what follows is that $\op{T}^{(S)}_{Kq}$ can be related to the
$K$th power of the generators (\ref{Stokop}), so they are intimately
linked to the moments of the Stokes variables, precisely our main
objective in this work. In particular, the monopole
$\varrho_{00}^{(S)}$ being proportional to the identity, is always
trivial, while the dipole $\varrho_{1q}^{(S)}$ is the first-order
moment of $\op{\mathbf{S}}$ and thus gives the classical picture, in
which the state is represented by its average value.

The complete characterization of the state demands the knowledge of
all the multipoles.  This implies measuring the probability
distribution of $\op{\mathbf{S}}$ in all the directions, and then
performing an integral inversion (put in another way, a whole
tomography), which turns out to be a hard
task~\cite{Karassiov:2004xw,Marquardt:2007bh,Muller:2012ys}. However,
in most realistic cases, only a finite number of multipoles are needed
and then the reconstruction of the $K$th multipole entails measuring
along just $2K+1$ independent
directions~\cite{Newton:1968ve,Klimov:2012ly}.

\section{Polarization quasidistributions}

The discussion thus far suggests that polarization must be specified by
a probability distribution of polarization states. As a matter of fact,
such a probabilistic description is unavoidable in quantum optics from
the very beginning, since $\{ \op{S}_{k} \}$ do not commute and thus
no state can have a definite value of all them simultaneously.

The SU(2) symmetry inherent in the polarization structure, as
discussed in the previous sections, allows us
to take advantage of the pioneering work of
Stratonovich~\cite{Stratonovich:1956qc} and
Berezin~\cite{Berezin:1975mw}, who worked out quasiprobability
distributions on the sphere satisfying all the pertinent
requirements.  This construction was later generalized by
others~\cite{Agarwal:1981bd,Brif:1998if,Heiss:2000kc,
Klimov:2000zv,Klimov:2008yb} and has proved to be very useful in
visualizing properties of spinlike systems~\cite{Dowling:1994sw, 
Atakishiyev:1998pr,Chumakov:1999sj,Chumakov:2000le,Klimov:2002cr}.

For each partial $\op{\varrho}^{(S)}$, one can define the SU(2) $Q$ function
as
\begin{equation}
  \label{eq:QSU2j}
  Q^{(S)} (\theta, \phi) = \langle S; \theta, \phi |
  \op{\varrho}^{(S)}  | S; \theta, \phi  \rangle \, ,
\end{equation}
where $ | S; \theta, \phi \rangle $ are the SU(2) coherent states
(also known as spin or atomic coherent states), given
by~\cite{Arecchi:1972zr,Perelomov:1986ly}
\begin{equation}
  \label{eq:defCS}
  |S; \theta , \phi \rangle = \op{D} (\theta, \phi ) 
  |S, -S \rangle \,  .
\end{equation}
Here $\op{D} (\theta, \phi ) = \exp (\xi \op{S}_{+} - \xi^{\ast}
\op{S}_{-})$ [with $\xi = (\theta / 2) \exp (- i \phi)$ and $(\theta,
\phi)$ being spherical angular coordinates] plays the role of a
displacement on the Poincar\'e sphere of radius $S$. The ladder
operators $\op{S}_{\pm} = \op{S}_{1} \pm i \op{S}_{2}$ select the
fiducial state $|S, -S \rangle$ as usual: $\op{S}_{-} |S , - S \rangle
= 0$. As we can appreciate, both the definition of the $Q$ function
and  the coherent states for SU(2) closely mimic their standard
counterparts for position-momentum. 

While for spins, $S$ is typically a fixed number, in quantum optics
most of the states involve a full polarization sector as in
equation~(\ref{eq:decrj}) and for the total polarization matrix
$\op{\varrho}$ we have
\begin{equation}
  \label{eq:QSU2}
  Q (\theta ,\phi ) = \sum_{S} \frac{2S+1}{4 \pi}
  Q^{(S)} (\theta ,\phi ) \, .
\end{equation}
The sum extends over the subspaces contributing to the state.  Since
the SU(2) coherent states are eigenstates of the total number operator
$\op{N}$, the sum over $S$ in (\ref{eq:QSU2}) attempts to remove the
total intensity of the field in such a way that $Q ( \theta, \phi)$
contains only relevant polarization information.  Furthermore, since
$|S; \theta , \phi \rangle$ are the only states saturating the
uncertainty relation~(\ref{eq:ursu2}), the definition of $Q (\theta,
\phi)$ is quite appealing, for it appears as the projection on the
states having the most definite polarization allowed by the quantum
theory.

On the other hand, as $Q$ contains the  whole information about the
state, its knowledge is tantamount to determining all the state
multipoles. Actually, the $Q$ function (and, more generally, any
$s$-parametrized quasidistribution) can be also written in
terms of $\varrho_{Kq}^{(S)}$~\cite{Agarwal:1981bh}:
\begin{equation}
  \label{eq:QSU2rj}
  Q^{(S)} (\theta ,\phi ) = \frac{\sqrt{4 \pi}}{\sqrt{2S+1}} 
\sum_{K=0}^{2S} \sum_{q=-K}^{K} C_{SS,K0}^{SS} \, \varrho_{Kq}^{(S)} \,
  Y_{Kq}^{\ast} (\theta, \phi) \, , 
\end{equation}
with $Y_{Kq} (\theta, \phi)$ being the spherical harmonics, which
constitute a complete set of orthonormal functions on the sphere.
This definition can be shown to be fully equivalent to
(\ref{eq:QSU2j}). Note also that the Clebsch-Gordan coefficient
$C_{SS,K0}^{SS}$ has a very simple analytical
form~\cite{Varshalovich:1988ct}:
\begin{equation}
  \label{eq:Cesp}
  C_{SS,K0}^{SS} = \frac{\sqrt{2S+1} (2S)!}
{\sqrt{(2S-K)! \, (2S+1 + K)!}} \, .
\end{equation}

By plugging (\ref{eq:QSU2rj}) in the general definition
(\ref{eq:QSU2}), we can express the $Q$ function as a sum over
multipoles:
\begin{equation}
  \label{eq:QsumK}
  Q (\theta, \phi) = \sum_{K=0}^{\infty} 
  Q_{K} (\theta, \phi) \, ,
\end{equation}
where each partial component can be written as
\begin{equation}
  \label{eq:QSU2K}
  Q_{K} (\theta, \phi) =
  \sum_{S=\lfloor K/2 \rfloor}^{\infty}  \sqrt{\frac{2S+1}{4 \pi}} 
  \sum_{q=-K}^{K} C_{SS,K0}^{SS} \, \varrho_{Kq}^{(S)} \, 
  Y_{Kq}^{\ast} (\theta , \phi )  \, .
\end{equation}
Here, the floor function $\lfloor x \rfloor$ is the largest integer
not greater than $x$. For the particular case of single $S$ (fixed
number of photons), the sum over $S$ has to be ignored.

The partial components $Q_{K}$ inherit the properties of $Q$, but they
contain exclusively the relevant information of the $K$th moment of
the Stokes variables. So, (\ref{eq:QsumK}) appears as an optimum tool
to arrange the successive moments and thus achieves our goals in this
paper.

%%%%%%%%%%%%%%%%%%%%%%%%%%%%%%%%%%%%%%%%%%%%%%%%%
\begin{figure}[t]
  \includegraphics[width=0.80\columnwidth]{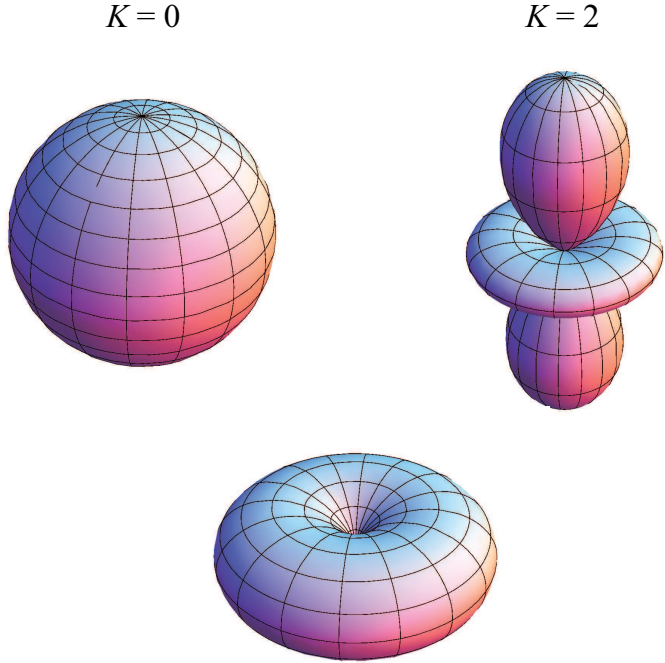}
  \caption{Plot of the monopolar ($K=0$) and quadrupolar ($K=2$)
    components of the state $|1, 0 \rangle$ (top), as well as the
    total $Q$ function (bottom). The dipolar component ($K=1$)
    vanishes, so this state lacks any first-orden information, which
    translates in the presence of \textit{hidden polarization}.}
  \label{Fig:NOONplots}
\end{figure}
%%%%%%%%%%%%%%%%%%%%%%%%%%%%%%%%%%%%%%%%%%%%%%%%%%

Let us illustrate our viewpoint with the simple example of
the state $| 1_{\mathrm{H}}, 1_{\mathrm{V}} \rangle$,  produced in
parametric down conversion. In the $|S, m\rangle$
notation, the state is $|1, 0 \rangle$ and its  $Q$ function can be
easily computed  according
to (\ref{eq:QsumK}) and (\ref{eq:QSU2K});
the final result is 
\begin{equation}
  \label{eq:Qt11}
  Q (\theta, \phi) = \frac{3}{4}
 \sqrt{\frac{1}{3 \pi}} \sin^{2} \theta  \, .
\end{equation}
It does not depend on $\phi$ and its shape is an equatorial belly,
revealing that the state is highly delocalized.  The partial
components are
\begin{equation}
  \label{eq:Qp11}
\fl
  Q_{0} (\theta, \phi) = \frac{1}{2}
 \sqrt{\frac{1}{3 \pi}}  \, ,
\quad
  Q_{1} (\theta, \phi) = 0 \, ,
\quad 
 Q_{2} (\theta, \phi) = - \frac{1}{2}
 \sqrt{\frac{1}{3 \pi}} \left ( \frac{3}{2} \cos^{2} \theta -
 \frac{1}{2} \right )  \, .
\end{equation}
The sum of these three terms gives, of course, the result
(\ref{eq:Qt11}), but anyway there is more information encoded in
(\ref{eq:Qp11}): the dipole contribution is absent, confirming that
this state conveys no first-order information (i.e., is
\textit{unpolarized} to that order). This is the reason why this is
the first state in which hidden polarization was detected. Figure 1
shows the partial $Q_{K}$ functions for this state, as well the global
one.

\section{Assessing higher-order polarization moments}

Let us consider the following quantity
\begin{equation}
  \label{eq:Q2}
  \mathcal{A} = \int  d\Omega \, 
  [ Q (\theta, \phi )   ]^{2}  \, ,
\end{equation}
where the integral extends over the whole sphere and $d\Omega= \sin
\theta d\theta d\phi$ is the solid angle. This function can be
interpreted as the effective area where the $Q$ function is different
from zero.  Similar proposals have already been used as measures of
localization and uncertainty in different
contexts~\cite{Heller:1987dq,
  Maassen:1988cr,Anderson:1993nx,Hall:1999oq,Gnutzmann:2001kl,
  Munoz:2012tg}. In polarization, (\ref{eq:Q2}) has been also used as
an essential ingredient in formalizing an alternative degree 
arising as the distance between the state's $Q$ function
and the $Q$ function for unpolarized light~\cite{Luis:2002ul}.

One might think  the use of the Wigner function preferable as a measure
of the area occupied by a quantum state in phase space. However, for
SU(2) $\int d\Omega [W (\theta, \phi)]^{2}$ takes exactly the same
value for all pure states, so that this provides a measure of purity
of quantum states rather than a measure of polarization. For this
compelling reason, we have instead employed the $Q$ function so far.

Note that (\ref{eq:Q2}) is invariant under SU(2) transformations:
this means that such an effective area  depends on the form of the $Q$
function, but not on its position or orientation on the  Poincar\'e
sphere.

Of course, the decomposition in multipoles (\ref{eq:QsumK}) is of
straightforward application here. Consequently, we can define the
magnitude
\begin{equation}
  \label{eq:Q2K}
  \mathcal{A}_{K} = \int  d\Omega \, 
  [ Q_{K} (\theta, \phi )   ]^{2}  \, ,
\end{equation}
with an analogous interpretation to that $\mathcal{A}$, but restricted
to the $K$th multipole.  Let us restrict ourselves to a fixed $S$ (and
drop the corresponding superscript for clarity); the generalization
for a sum of $S$s is direct. When the explicit form of $Q_{K}$ in
(\ref{eq:QSU2K}) is used, $\mathcal{A}_{K}$ reduces to
\begin{equation}
  \label{eq:AK}
  \mathcal{A}_{K} =   \sum_{S=\lfloor K/2 \rfloor}^{\infty}
  \frac{2S+1}{4 \pi}  \sum_{q=-K}^{K}
  \left ( C_{SS,K0}^{SS}  \right )^{2}
  | \varrho_{Kq}^{(S)}|^{2} \, .
\end{equation}
In this way, $ \mathcal{A}_{K} $ can be reinterpreted as a measure of
the strength of the corresponding multipole, confirming that it
provides full information about the state $K$th moment.

%%%%%%%%%%%%%%%%%%%%%%%%%%%%%%%%%%%%%%%%%%%%%%%%%
\begin{figure}
 \includegraphics[width=0.60\columnwidth]{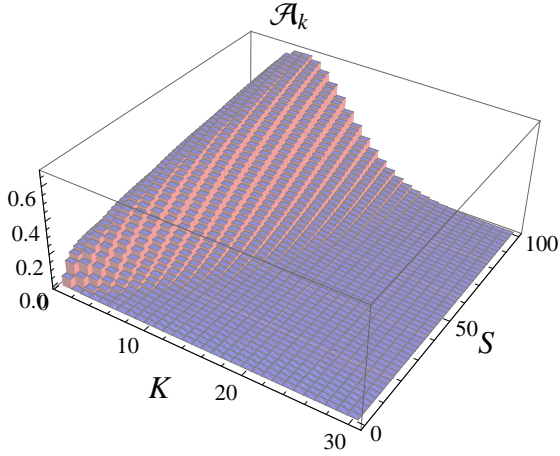}
  \caption{Function $\mathcal{A}_{K}$ as a function of the multipole
    order $K$ for a SU(2) coherent state for different values of $S$.}
\end{figure}
%%%%%%%%%%%%%%%%%%%%%%%%%%%%%%%%%%%%%%%%%%%%%%%%%%

As an appealing illustration of our method, we analyze the outstanding example of SU(2) coherent states. 
Without lack of generality,  we  deal with the south pole
$ | S,  - S \rangle$, since from (\ref{eq:defCS})  any other coherent state can
be obtained by the application of a displacement to that state. 
The  associated multipoles turn out to be $\varrho_{Kq}^{(S)}
= \sqrt{(2K+1)/(2S+1)} C_{SS,K0}^{SS}$, so that
\begin{equation}
  \label{eq:AKCS}
  \mathcal{A}_{K} = \frac{2K+1}{4 \pi} 
\left ( C_{SS,K0}^{SS}  \right )^{4} \, .
\end{equation}
In figure~2 we have plotted $\mathcal{A}_{K}$ as a function of $K$ and
$S$. The first multipoles contribute always the most to the
state localization, something that one could expect from physical
intuition. However, as $S$ gets larger, the number of multipoles to
take into account also increases.

\section{Concluding remarks}

Multipolar expansions are a commonplace and a powerful tool in many
branches of physics.  We have applied such an expansion to the
polarization density matrix, showing how the corresponding state
multipoles represent higher-order correlations in the Stokes
variables. This paves the way to a systematic characterization of
quantum polarization fluctuations that, paradoxically, is still
missing in the realm of quantum optics. Such a complete programme is
presently in progress in our group.

\ack

We thank G~S~Agarwal, G~Bj\"{o}rk, J~H~Eberly, W~P~Schleich,
T~H~Seligman and K~B~Wolf for useful discussions.  Financial support
from the EU FP7 (Grant Q-ESSENCE), the Spanish DGI (Grant
FIS2011-26786), the UCM-BSCH program (Grant GR-920992) and the Mexican
CONACyT (Grant 106525) is acknowledged.

\newpage

%\bibliographystyle{iopart-num} 
%\bibliography{Polarization}

\providecommand{\newblock}{}

\end{document}